\documentclass[%
 reprint,
 superscriptaddress,
nofootinbib,
 amsmath,amssymb,
 aps,
superscriptaddress
]{revtex4-2}

\usepackage{graphicx}% Include figure files
\usepackage{dcolumn}% Align table columns on decimal point
\usepackage{bm}% bold math
\usepackage{hyperref}% add hypertext capabilities
\usepackage[mathlines]{lineno}% Enable numbering of text and display math
\usepackage{diagbox}
\usepackage{tabularx}
\usepackage[dvipsnames]{xcolor}
\usepackage{adjustbox}

\begin{document}

\preprint{APS/123-QED}

\title{CURE-like, not cure-all: Varying broad relevance in experimentation labs produces similar student outcomes}

% \title{How relevant is relevance? CURE-like physics labs show similar student outcomes to other experimentation labs}

\author{Elly Markert}
\affiliation{Department of Physics, Cornell University, Ithaca, New York 14853, USA}

\author{Mike Verostek}
\affiliation{Laboratory of Atomic and Solid State Physics, Cornell University, Ithaca, New York 14853, USA}
\affiliation{Laboratory of Elementary Particle Physics, Cornell University, Ithaca, New York 14853, USA}

\author{Matthew Dew}
\affiliation{Laboratory of Atomic and Solid State Physics, Cornell University, Ithaca, New York 14853, USA}

\author{Xuan Chen}
\affiliation{Laboratory of Elementary Particle Physics, Cornell University, Ithaca, New York 14853, USA}
\affiliation{Department of Physics, University of Toronto, Toronto, Ontario, M5S 1A7, Canada}

\author{Mark Lory-Moran}
\affiliation{Department of Physics, Cornell University, Ithaca, New York 14853, USA}

\author{Sofi Padavan}
\affiliation{Department of Physics, Cornell University, Ithaca, New York 14853, USA}

\author{Hissaa Walia}
\affiliation{Department of Physics, Cornell University, Ithaca, New York 14853, USA}

\author{Peter Wittich}
\affiliation{Laboratory of Elementary Particle Physics, Cornell University, Ithaca, New York 14853, USA}

\author{N.G. Holmes}
\email{ngholmes@cornell.edu}
\affiliation{Laboratory of Atomic and Solid State Physics, Cornell University, Ithaca, New York 14853, USA}

\date{\today}

\begin{abstract}

Physics labs that engage students in practices authentic to experimental physics (experimentation-based labs) are being implemented to modernize the undergraduate physics curriculum and broaden participation in physics. Accordingly, prior research has positioned Course-Based Undergraduate Research Experiences (CUREs) as a means to extend the benefits of authentic undergraduate research experiences to more students. However, CUREs are resource-intensive and difficult to implement; a continuous stream of novel research projects adaptable for undergraduate courses is rare. Further, little is known about which specific components of a CURE are crucial to improving student outcomes and which components could be scaled back to improve feasibility for a wider range of class settings. In this study, we aim to isolate the component of broad relevance by running two experimentation-based labs in parallel: one ``CURE-like'' that increases broad relevance through the use of muon detectors, and one that uses equipment typical to an introductory physics lab and not relevant beyond the classroom. We measure student outcomes for both experimental critical thinking skills and attitudes towards physics labs. We use hierarchical linear modeling to compare student outcomes between the two labs. We find that both experimentation-based labs produce similar student outcomes. Our results suggest that increased levels of broad relevance may not inherently improve gains in student learning or attitudes. Future work should further investigate which components of different experimentation-based lab formats are associated with gains in student outcomes. Although this study did not implement a full CURE, our findings align with a growing body of evidence challenging the idea that CUREs are uniquely positioned to achieve superior student outcomes over other well-designed experimentation-based labs.

%just putting some stuff here: 
% Why are we doing this/what’s been done, where there’s a gap:
% - Prior research has proposed CUREs as a very beneficial way to improve student experimental skills
% - But there has not been a lot of work done in physics specifically and more so CUREs are like really hard to implement. Resource intesive and also hard to have ideas that r truly CUREs
% - So which parts r most important? A lot of labs have the first 3 aspects, theres not a lot of research that isolates aspects of CUREs, esp broad relevance 
% How we are filling this gap:
% - We run class that has two flavors of lab and one of them is CURE like w/ increased levels of broad relevance and the other one isn’t 
% Methods and what we’re doing with it:
% - We take data using the PLIC which tests lab critical thinking skills and also physics attitudes
% - We use an HLM to find the effect of the CURE-like class on these things
% - Then we compare them
% What did we find:
% - Zero significant difference between the CURE-like and skills based class, in any outcome ever
% - But both labs still show gains?
% What are the implications of what we found:
% - Maybe broad relevance is not that important
% - Maybe gains in a good skills based lab = a CURE/like lab
% - we need to do more stuff

\end{abstract}

\maketitle

\section{\label{sec:Intro}Introduction}

Transforming introductory physics labs to engage students in the skills, practices, and ways of thinking authentic to experimental physics has been identified as a critical priority for physics educators \cite{noauthor_engage_2012_v2, quinn_framework_2012, subcommittee_of_the_aapt_committee_on_laboratories_aapt_2014, heron_phys21_2016, laursen_levers_2019, singer_americas_2005, holmes_introductory_2018, smith_best_2021}. Introductory lab curricula are often dichotomized into ``traditional'' labs, characterized by prescribed procedures and verification of physical concepts for content reinforcement, and ``non-traditional'' labs that eschew such rigid structure to focus on students' development of experimental skills. Referring to non-traditional lab formats collectively as ``experimentation-based'' labs is one way to articulate this dichotomy, and much effort in physics education research has already gone toward designing such labs to better immerse students in the skills, practices, and ways of thinking aligned with experimental physics \cite[\textit{e.g.},][]{may_historical_2023, etkina_role_2002, beichner_introduction_2000_v2, dounas-frazer_modelling_2018, zwickl_incorporating_2014, smith_direct_2020}. 

In biology, many instructors have responded to similar calls by replacing traditional labs with Course-Based Undergraduate Research Experiences (CUREs), which are designed to expand students' access to authentic research experiences \cite{auchincloss_assessment_2014, rodenbusch_early_2016, bangera_course-based_2014}. The primary difference between CUREs and other common experimentation-based labs is that students in a CURE perform novel, and often publishable, scientific research. CUREs provide a larger and more diverse range of students with the opportunity to participate in research, which studies suggest could help support broader persistence in science \cite{werth_impacts_2022, bangera_course-based_2014, estrada2016improving}. 

Despite strong evidence that CUREs are beneficial for students, their implementation in physics remains rare \cite{buchanan_current_2022}. CUREs are defined as having five key components: 1) use of scientific practices, 2) discovery, 3) broad relevance, 4) collaboration, 5) iteration \cite{auchincloss_assessment_2014}. The second and third of these requirements most strongly distinguish CUREs from other experimentation-focused labs: in a CURE, students' work should 1) result in novel discovery and 2) have ``broad relevance,'' meaning that their results are of interest to the wider scientific community \cite{auchincloss_assessment_2014}. Both of these requirements may be satisfied by engaging students in publishable scientific research; however, this creates a significant burden on instructors to sustain a pipeline of innovative research projects across semesters. The logistical challenges associated with implementing CUREs in physics raise a central question for educators seeking to promote laboratory reform: which components of the CURE model are necessary to support student outcomes, and which may be adapted or constrained to improve scalability? If the quality of student outcomes can be maintained while engaging students in work that is highly contextualized in contemporary physics, but whose results are not directly relevant to researchers, it would substantially lower barriers to implementing such a ``CURE-like'' course in physics. Reviews of CURE research \cite{buchanan_current_2022, corwin_modeling_2015}, however, indicate that little research to date has attempted to isolate the effects of individual CURE components on student outcomes. Whether a modified ``CURE-like'' course produces better student outcomes than existing experimentation-based labs, or perhaps matches the outcomes of a standard CURE, remains to be seen.

This paper begins to address this gap by systematically evaluating the extent to which labs with a similar focus on scientific practices, discovery, iteration, and collaboration, but a different focus on broad relevance, produce similar benefits to all students. Specifically, we seek to answer the following research questions: 

\begin{enumerate}
    \item How does varying the broad relevance of lab content relate to student outcomes, particularly lab critical thinking skills and student attitudes?
    \item Do we observe evidence of differential outcomes across demographic groups in response to varying the relevance of lab content? 
\end{enumerate}

To answer these questions, we present results from a quasi-experimental study contrasting two labs: an experimentation-based lab in which the concepts and apparatus are at the level of introductory physics and not relevant beyond the classroom, and a CURE-like lab in which the concepts are in line with cutting-edge experimental particle physics research and applicable, though not directly relevant, to particle physics research. We find no statistically significant difference in postsurvey scores between students in the two lab conditions on five measured constructs: experimental critical thinking skills, self-efficacy, perceived agency, belonging, and recognition.

% We begin in Section \ref{sec:Background} by briefly reviewing relevant literature on physics identity to illustrate why answering these questions is important for supporting broader persistence in science. We then situate our study within prior CURE literature, highlighting several studies indicating that students develop many of the important research skills and understandings from labs that \textit{simulate} authentic research as those that directly engage students in publishable research. The two lab conditions studied in this paper are detailed in Section \ref{sec:Methods}, and results are reported in Section \ref{sec:Results}. Implications for instruction and discussion of future work are offered in Section \ref{sec:Discussion}. 

\section{\label{sec:Background}Background}

\subsection{Instructional labs to support a modern curriculum and broad participation in physics}

National calls to reform the undergraduate physics curriculum have increasingly highlighted the need to prepare students for the modern STEM workforce \cite{heron_phys21_2016, nsf2022strategic, el2026experimental}. Despite significant changes to \textit{how} students are being taught physics \cite{meltzer_brief_2015}, the content of the undergraduate curriculum has remained largely static for over 50 years \cite{noauthor_adapting_2013}. It remains predominantly organized around theoretical instruction, with engagement in experimental practices typically confined to time spent in one to two lab courses a year \cite{cardona2021access}. Within the lab sessions that do exist, students rarely encounter the open-ended, iterative, and decision-driven practices that characterize real scientific work \cite{holmes_introductory_2018}. Rather, many labs remain rigidly structured and focused on verifying known results from lecture.  Indeed, physics graduates commonly report that they wish they had learned more experimental skills during their undergraduate training, including hands-on time spent designing, building, and troubleshooting real equipment \cite{heron_phys21_2016}. To better align the curriculum with relevant postgraduation outcomes, researchers and educators must work to develop and evaluate classroom experiences that support understanding of how science is done in ``real-world'' settings.

Re-imagining the undergraduate physics curriculum can also better support the success and retention of a broader, more diverse population of physics students \cite{nsf2022strategic, noauthor_adapting_2013, noauthor_engage_2012_v2}. Physics remains one of the least diverse STEM disciplines with regard to participation by women and other historically underrepresented groups \cite{aps2024degrees, ipeds}. Although students choose to major in physics for many reasons, education researchers have demonstrated the major role that physics identity plays in students' decision to pursue the field \cite{hazari_connecting_2010, carlone_understanding_2007, lock2013physics}. Students form identities within their ``figured worlds'' of physics---the culturally and socially constructed spaces of meaning that shape how students understand the discipline and their possible roles within it \cite{lock_discussing_2016, holland2001identity}. Students' figured worlds are continuously constructed and re-constructed based on experiences both in and out of the classroom, which collectively shape their perceptions of what physics is and who participates in it. 

For students from historically underrepresented groups, the narratives that have traditionally dominated physics classrooms and culture rarely exemplify models of participation in the discipline that reflect their own backgrounds \cite{barton_we_2010, gonsalves_exploring_2018}. Rather, students most commonly encounter exemplars of physicists whose trajectories are framed as the inevitable result of innate genius and talent, which constrains the space of plausible pathways that students feel capable of pursuing. 
% Students generally perceive physicists as more innately brilliant, more socially awkward, and less collaborative than other scientists \cite{leslie_expectations_2015, bruun_identifying_2018}, which contrasts with the reality of contemporary physics research conducted by diverse and highly collaborative teams who grapple with complex problems over many years.
With more opportunities to engage in practices that mirror what practicing physicists do, students may develop a more informed and complete figured world of physics and navigate how they fit within it.  

One of the most common ways for students to gain exposure to the work of professional physicists is through participation in undergraduate research, which has been shown to enhance students' physics content knowledge as well as their sense of belonging, self-efficacy, and physics identity \cite{lopatto_survey_2004, zohrabi_alaee_impact_2022, hunter_becoming_2007}. Being part of a research group can expand students' perceptions of who participates in physics and what that participation looks like. Several studies indicate that undergraduate research experiences are particularly powerful for racially underrepresented and first-generation students \cite{kuh2008high, finley2013assessing, eagan2013making}. Access to these experiences, however, remains uneven, with these same students less likely to participate \cite{kuh2008high, hu2025bridging, pierszalowski2020research}. This misalignment between impact and access highlights the critical need for development of scalable and inclusive classroom experiences that better represent what physicists actually do.

Instructional physics labs are well positioned to simultaneously address the skill gap in the undergraduate curriculum and the urgent need to broaden the base of students participating in physics. Labs are one of the few places in the curriculum designated for engaging students in experimental practices. Numerous studies illustrate that experimentation-based lab formats are more effective than traditional, concept-reinforcement labs for supporting student outcomes related to experimental skills and attitudes \cite{etkina_how_2008, walsh_assessment_2018, zwickl_model-based_2015, sulaiman2023students, wilcox_developing_2017}, and that the benefits of these labs extend to all students \cite{walsh_skills-focused_2022}.  

Despite the significant gains in student achievement due to such curricular reforms, however, these models for introductory labs remain fundamentally limited in their capacity to engage students in what many would consider ``real'' scientific research. The physics of pendulums, ramps, and circuits is well established, so the experiments conducted in introductory labs generally do not generate novel results or reflect the kinds of experiments conducted by practicing physicists. This fact has led researchers and instructors to explore whether more authentic lab experiences that encourage students to probe open questions and contribute to novel knowledge production can be integrated into the introductory curriculum.

\subsection{The promise and practical challenges of CUREs for engaging students in physics research}

Course-Based Undergraduate Research Experiences are an increasingly common model for replacing STEM labs that aim to address the gap between traditional introductory lab experiences and authentic scientific research \cite{auchincloss_assessment_2014, brownell_toward_2015_v2, rodenbusch_early_2016, bangera_course-based_2014}. In a CURE, students conduct novel research projects that seek to answer questions whose answers are unknown in the field~\cite{auchincloss_assessment_2014}. Relevant CURE literature suggests that students in CUREs may achieve many of the outcomes traditionally obtained through undergraduate research experiences \cite{corwin_modeling_2015, dolan2016course, krim_models_2019, werth_impacts_2022}, but because students enroll in CUREs as part of their normal class schedule, the barrier to participating in research is significantly lowered.

Broadly, CUREs are defined by five principles \cite{auchincloss_assessment_2014}:
    \begin{enumerate}
        \item Use of scientific practices: Students ask questions, build and evaluate models, make hypotheses, design studies, choose methods, use tools, gather and analyze data, etc. 
        \item Discovery: Students address ``novel scientific questions'' that have not been answered by the student or the instructor.
        \item Broadly relevant or important work: Students perform work with ``impact and action beyond the classroom'' that is relevant to external stakeholders
        \item Collaboration: Students work together to tackle problems, improve work through peer feedback, and develop teamwork skills.
        \item Iteration: Students repeat and revise their own work, revise aspects of other students' investigations in the course, and/or revise work that has been done across successive offerings of the same course.
    \end{enumerate}
While students in other non-traditional, experimentation-based laboratory courses are likely to engage in scientific practices, collaboration, and iteration, the components that most strongly differentiate a CURE are the discovery and broad relevance components. These components require that the results of students' experiments are not already known by either students or instructors (discovery) \emph{and} hold value for the wider scientific community (broad relevance), which is not often the case for introductory physics lab activities.

Although CUREs have proliferated in undergraduate biology education over the past decade \cite{treibergs2025scoping}, only one large-scale CURE has been conducted and studied in the physics education literature \cite{werth_impacts_2022}, with other small-scale CUREs studied in the physics and astronomy education literature \cite{wooten_investigating_2018, hewitt_development_2023, rabosky_cure_2025}. The first large-scale physics CURE, the Colorado PHysics Laboratory Academic Research Effort (C-PHLARE), was conducted remotely during the COVID-19 pandemic and relied on analysis of a large open source data set to engage students in publishable research \cite{werth_impacts_2022}. Research indicated that students in the CURE gained research skills, expressed more interest in experimental physics, and felt like they engaged in real-world research during the course \cite{werth_assessing_2022, oliver_student_2023, werth_enhancing_2023, werth_impacts_2022}. 

C-PHLARE demonstrates that large-scale physics CUREs are possible and can produce successful student outcomes. Still, due to its focus on data analysis, C-PHLARE's structure does not address the feasibility of engaging students in a CURE that emphasizes hands-on experimental physics practices such as designing experiments and building and testing hardware. These practices are foundational to many areas of experimental physics and are also important for students to experience in lab to prepare them for the workforce~\cite{heron_phys21_2016}. However, an in-person, hands-on equipment-based CURE likely presents more significant barriers to implementation. Designing brand new laboratory activities, investing in new equipment, and continuously developing novel research questions that remain viable for large enrollments may be too resource-intensive for many departments.

These challenges are directly tied to the discovery and broad relevance requirements that define the CURE model. Engaging students in the experimental practices of designing and building hardware is not inherently problematic---hands-on practice with lab equipment is a regular occurrence in laboratory courses. The challenge lies in contextualizing these activities to make them both novel and relevant beyond the classroom. This raises critical questions regarding the relative value added to students’ lab experiences due to the discovery and broad relevance components of the CURE. Relaxing these requirements would substantially lower barriers to implementing a ``CURE-like'' course that engages students in more broadly relevant work than is typical in introductory physics labs, but does not produce fully novel research. It remains unclear, however, whether this would yield student outcomes comparable to those of a standard CURE, or how those outcomes would compare to other experimentation-based lab formats. %Indeed, prior studies comparing traditional CUREs to skills-based labs in order to isolate the effects of relevance and discovery have shown mixed results.  

% To better understand what is gained or lost when certain CURE components are scaled back, the current study compares student outcomes in a modified CURE-like course to student outcomes in an existing skills-based lab. To help situate our study within current CURE literature, we next review several prior studies that have attempted to isolate the effects of relevance and discovery on student outcomes, showing mixed results. 

%More broadly, the transition to a CURE model that fulfills the discovery and broad relevance components may pose insurmountable barriers for some departments due to the level of resources requir Indeed, more research is needed to better understand which outcomes are associated with which CURE components \cite{corwin_laboratory_2015, buchanan_current_2022}.ed to sustain it. 

\subsection{Rethinking the impact of discovery and broad relevance on student outcomes}

Research on CUREs characterizes discovery and broad relevance as critical elements of the experience that offer benefits to students beyond other laboratory formats \cite{corwin2018need, corwin_effects_2018, cooper2019impact}. In a qualitative comparison of students in a CURE and an experimentation-based lab, \citeauthor{goodwin_is_2021} found that failure, iteration, and relevant discovery acted in conjunction to enhance how CURE students perceived the authenticity of their experience \cite{goodwin_is_2021}. Another study in the context of chemistry education compared students' gains in understanding of the nature of science across traditional, experimentation-based, and CURE (or research-based) lab courses. The students in the CUREs were found to have more sophisticated conceptions of experiments and theories than their counterparts, and were also more likely to discuss their personal experiences in the course during interviews \cite{russell_comparative_2011}. Hence, these studies support an integral role for discovery and broad relevance. 

In contrast, other research has suggested that students develop many of the same important research skills and understandings from labs that \textit{simulate} authentic research \cite{rowland_we_2016, ballen_discovery_2018, hebert2021open, lansverk2020comparing}. For example, \citeauthor{rowland_we_2016} compared student reflections from a CURE with reflections from an experimentation-based laboratory that focused on experimental skills and discovery but not generating novel research outcomes. The authors found the content of the reflections to be similar across the lab types and concluded that students shared the perception that their experiences ``provided them with insight into how a laboratory runs, how research is done, and what scientists do in their everyday jobs'' \cite{rowland_we_2016}. The authors specifically noted the ``surprising'' number of students in the experimentation-based lab course who discussed feeling like it was representative of a real research lab. This phenomenon is not isolated. In a meta-analysis of 24 studies that used a common survey measuring students’ perceived engagement in discovery- and relevance-oriented activities, \citeauthor{beck2023can} found substantial overlap across lab types. In particular, students in CUREs did not report meaningfully different levels of discovery or relevance compared to students in other experimentation-based laboratory courses \cite{beck2023can}. 

Collectively, these studies leave open the possibility that similar student outcomes may be achievable in experimentation-based labs as in CUREs (or CURE-like courses) by maintaining some elements of discovery while constraining broad relevance. Experiments built around research questions that are unknown to students, even if their answers are well established in the physics literature, may provide them with many of the same benefits as conducting novel research. Prior studies evaluating undergraduate research experiences in physics lends some support to this idea as well, finding that the pursuit of publishable research questions was not an important component driving students' perceptions of benefits from undergraduate research \cite{holmes_examining_2016}. Meanwhile, a longitudinal study of physics students’ REU experiences by \citeauthor{zohrabi_alaee_impact_2022} found that engaging in the specific practices associated with novel research did not, on their own, necessarily lead to positive student outcomes. Instead, students’ participation in novel research fostered a sense of meaningful contribution to their research groups, which in turn strengthened their physics identity \cite{zohrabi_alaee_impact_2022}. It is unclear how the entanglement of novel research and social recognition within a research group would translate to a classroom setting. 

Ultimately, it is clear that more work is needed to better elucidate the roles of discovery and broad relevance in introductory labs, particularly in physics. 
%No work that we are aware of has systematically compared how labs across the continuum of discovery and broad relevance affect student outcomes. Moreover, little of the cited research has been conducted in the context of physics. 
Filling this gap in the literature is critical for understanding how to design readily scalable lab experiences that support a broader base of students participating in physics. Hence, we have designed the current study to contrast two conditions of labs that vary systematically in broad relevance in order to characterize their effects on student outcomes.

\subsection{Two flavors of lab to systematically vary broad relevance: SALT and PEPPER}

\begin{figure}
    \centering
    \includegraphics[]{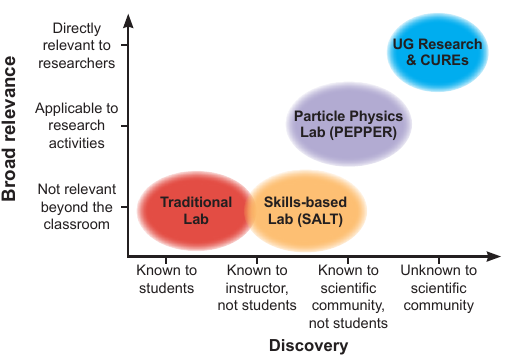}
    \caption{Representation of the range of discovery and broad relevance across different physics lab types. Broad relevance varies on a continuum according to the lab's relevance to the physics community, while discovery varies from exploring questions whose answers are known to everyone in the class to answers unknown to the scientific community. Blurred edges of the lab types exemplify that their boundaries on the diagram are not precise.}
    \label{fig:relevance_disc}
\end{figure}

In this study, we build on prior work by first disentangling and defining the range of discovery and broad relevance available to introductory physics students (see Fig. \ref{fig:relevance_disc}). Drawing on definitions in the CURE and inquiry-based learning literature, we conceptualize discovery according to who knows and does not know the answer to the question being investigated \cite{auchincloss_assessment_2014, ballen_discovery_2018, buck_research_2008}. We characterize discovery in introductory labs as a continuum ranging from exploring questions whose answers are known to everyone in the class (e.g., verification of theory) to answers unknown to the scientific community. Intermediate levels include exploring questions whose answers are known to the instructor but not the student (e.g., labs on topics not yet addressed in class), or whose answers are known or understood by the broader scientific community but not by the students or the instructor (e.g., labs on topics that involve collecting data about objects from home).

While this definition of discovery is straightforward, the meaning of ``relevance'' in science education is complex and multidimensional, with a review by \citeauthor{stuckey2013meaning} finding that it is often inadequately conceptualized in the literature \cite{stuckey2013meaning}. For the purposes of this study, we base our conceptualization of ``broad relevance'' on that of \citeauthor{auchincloss_assessment_2014}, who describe broadly relevant work as having ``impact and action beyond the classroom'' and as ``[fitting] into a broader scientific endeavor that has meaning beyond the particular course context'' \cite{auchincloss_assessment_2014}. The CURE literature commonly emphasizes the outward-facing nature of broad relevance. The relevance is to \textit{external stakeholders} \cite{corwin_modeling_2015, dolan2016course, buchanan_current_2022}. This stands in contrast to other dimensions of relevance such as ``personal relevance,'' which \citeauthor{kapon2018disciplinary} describe as students’ own sense of benefit, value, meaningfulness, and agency in the work \cite{kapon2018disciplinary}. \citeauthor{cooper2019impact} make this contrast explicit in their study on the impact of broad relevance, noting that ``the term `broadly' is used to distinguish between relevance beyond the course and personal relevance'' \cite{cooper2019impact}. While this framing does not preclude the possibility that a project may engage students in personal relevance while also being broadly relevant (\textit{e.g.}, a sense of contributing to a larger scientific endeavor may foster curiosity and meaningfulness), it clearly delineates broad relevance and personal relevance as distinct constructs.

This outward-facing characterization of broad relevance is not universal in the CURE literature. For instance, \citeauthor{brownell_toward_2015_v2} discuss broad relevance both in terms of interest to the research community as well as ``engaging students in authentic tasks of relevance to their own lives,'' thereby blending several dimensions of relevance \cite{brownell_toward_2015_v2}.  The authors also suggest that discovery and broad relevance be treated as one construct, since making novel discoveries should relate to increasing student excitement and engagement. \citeauthor{corwin_laboratory_2015} similarly argue that the two constructs are the same, finding that survey questions related to student perceptions of these constructs loaded onto a single factor \cite{corwin_laboratory_2015}. However, it is unclear that \textit{student perceptions} of broad relevance and actual relevance to external stakeholders, as emphasized in the definitions offered above, would necessarily align. Given this ambiguity, we adopt the externally facing definition described above, which most cleanly preserves the distinction from personal relevance.  

Operationally then, we characterize broad relevance as engaging students in work that has application \textit{and} relevance to physics research outside the classroom. We add application to emphasize that the students' results may not be directly relevant, but that the lab procedures, equipment, and concepts with which they are engaged may be applicable in contemporary physics research. As illustrated in Fig. \ref{fig:relevance_disc}, we view broad relevance as related to but distinct from discovery. Labs with low broad relevance would include those that are not relevant beyond the classroom (\textit{e.g.}, traditional labs about introductory-level physics concepts), while labs with high broad relevance would be directly relevant to researchers in the discipline (\textit{e.g.}, related to a faculty member's research program). Intermediary relevance could include a lab where the concepts and apparatus align with contemporary experimental research and are applicable to conducting research, though the experimental \emph{results} are not directly relevant to the physics community.

With these conceptualizations in mind, we designed and carried out a quasi-experimental study contrasting two flavors of labs that systematically vary along the axis of broad relevance. The first is what we will refer to as the SALT lab, where concepts and apparatus are at the level typical of introductory physics and are not relevant beyond the classroom. The second lab is what we will refer to as the Physics of PEPPER lab, where the concepts and apparatus are applicable to experimental particle physics research, though results are not directly relevant to particle physics researchers.\footnote{The PEPPER label was initially an acronym that combined ``Experimental Particle Physics'' and ``Physics Education Research.'' The SALT label was assigned as a convenient contrast. These labels are not meaningful acronyms and should not be considered as ``named teaching methods''~\cite{sundstrom_relative_2026, strubbe_beyond_2020}.} We do not assume that students will find muons or the physics behind them personally meaningful, but we do claim that the tasks and equipment involved in PEPPER are unambiguously more relevant to contemporary physics research than those in SALT. Meanwhile, the two conditions of labs have similar levels of discovery, involving questions where the answer is unknown to the students and often to the instructor, but not at a publishable level. Our conditions therefore evaluate the extent to which labs that include similar focus on scientific practices, discovery, iteration, and collaboration---but a different focus on broad relevance---produce similar benefits to students.

\section{\label{sec:Methods}Methods}
\subsection{Course Context}
Our study focuses on a large introductory experimental physics course primarily taken by physics and engineering students. Data were collected over a period of two semesters, referred to as semester 1 and semester 2. Instruction was delivered entirely in person. The learning outcomes shared for both the SALT and PEPPER conditions of the course were as follows (see Ref.~\cite{holmes_operationalizing_2019}):

\begin{enumerate}
    \item[1.] Collect data and refine an experimental procedure iteratively and reflectively
    \item[2.] Evaluate the processes and outcomes of an experiment qualitatively and quantitatively
    \item[3.] Extend the scope of an investigation whether or not the results come out as expected
    \item[4.] Communicate the process and outcomes of an experiment
    \item[5.] Conduct an experiment collaboratively and ethically
\end{enumerate}

Each learning outcome included several sub-outcomes, ranging from comparing measurements to sharing responsibility for experimental tasks. Students attended one two-hour lab section (with 20-25 students per section) and one 50-minute lecture per week (with 200-300 students per lecture). All students attended one of two weekly lecture sections, with no divisions for SALT or PEPPER lab conditions. Students also completed approximately weekly homework tutorials or reflection exercises. 

The structure of group work was the same for both semesters and conditions. Students worked in groups of two to four for each experiment. Lab groups were self-selected in the first lab session and changed twice over the course of the semester. For the final lab module, lab section TAs could intervene in group arrangement. Each group also completed a ``partner agreement'' every time they were rearranged. In these agreements, members introduced themselves to the group, discussed collective and individual goals in the course, and planned out group organization such that each member could be involved in their desired lab tasks \cite{dew_group_2024}. 

%and  typically changed twice throughout the semester.

Lab groups each submitted one set of lab notes by the end of the lab period, which were graded by the teaching assistants (TAs). The GPA scale of zero to four (shared with students at the start of the semester): four (or A) meant that the notes were generally complete with appropriate reasoning; three (or B) meant that the notes had one or more flaws or areas for improvement on data collection, evaluation, interpretation, and/or communication; two (or C) meant there were major flaws, such as major limitations in the experimental design, uninterpretable data or figures, or data interpretation; one (or D) meant there were several such major flaws, and zero (or F) meant nothing was submitted. In semester 2, a ``check-out'' system was used; in each session, students prepared their final claims and evidence, which they could check-out in person with their TA for feedback, if time permitted, or they could submit to be graded after the lab session ended. Check-outs were graded again on a zero to four GPA scale, but focused on evaluating the claim and evidence for the claim: four (or A) meant the claim is clearly articulated, the evidence supports the claim, and the notes indicate the students collected the evidence themselves; three (or B) meant there were one more flaws or areas for improvement, such as the evidence was insufficient to support the claim, the evidence supports multiple claims, or the claim itself is insufficient; two (or C) meant there were major flaws, such as the claim does not align with the evidence, major limitations in the experimental design, uninterpretable data or figures; one (or D) meant there were several such major flaws, and zero (or F) meant nothing was submitted. Students who checked-out in person could make changes to their notes, experiment, or claims based on the TA feedback, if time permitted.

\subsection{Instructional conditions}
As described above, the course was divided into two conditions of lab sections: SALT and PEPPER.

The SALT labs, previously described in Refs. \cite{smith_direct_2020, kalender_restructuring_2021}, were designed as open-ended labs intended to promote experimentation-based learning rather than demonstrate or teach physics concepts. In the first nine weeks of the course, students were given a set of experiments and broad goals each week. In groups, they set up or designed their own apparatus, collecting and analyzing their data with the guidance of a TA (some lab sections also had undergraduate learning assistants for support). Many experimentation decisions were left up to the students, with increasing freedom over time~\cite{kalender_restructuring_2021}. In the final three weeks of the course, students proposed and investigated their own research question, culminating in a final presentation where groups shared their findings with each other in a slideshow (semester 1) or poster (semester 2) format. 

The experiments (versions of which are available at physport.org/curricula/thinkingcritically) involve springs, pendulums, circuits, and other equipment and phenomena common to an introductory physics sequence. Between semesters one and two, the order of experiments in the SALT lab was changed. In semester 1, all of the mechanics experiments took place before the electricity and magnetism (E\&M) experiments. In semester 2, one of the later E\&M experiments was moved towards the beginning of the course to promote inquiry and open-ended experimentation from the beginning of the course, and the remaining labs alternated between mechanics and E\&M experiments. The format of the presentations also changed between semesters. In semester 1, student groups gave 10 minute presentations to their peers about their projects. In semester 2, students presented single-slide, digital ``posters'' and took turns presenting and interacting with each other's posters. The rest of the curriculum was unchanged.

The PEPPER labs followed the same experimentation-based framework and learning goals as the SALT labs, but replaced the lab equipment used with a tabletop cosmic ray muon detector adapted from the MIT CosmicWatch \cite{axani_cosmicwatch_2018}. The course started with an introduction to particle physics, muons, and the detectors themselves. The subsequent four lab sessions involved experiments focused on testing technical aspects of the detectors: how do we quantify uncertainty and precisely measure count rate, how do changes to the scintillator and trigger threshold affect the measured count rate, and how do we take measurements in coincidence mode? The final five lab sessions were student-led projects, with similar structure to those in the SALT lab (albeit using the muon detectors). Students could leave the lab rooms with their detectors, so long as they returned with them by the end of the scheduled lab session. The course culminated in a final presentation session that used the same format as that for the SALT labs.

% To emphasize the broad relevance of their work in the lab and communicate to students the role of muons in experimental particle physics, presentations were given at the beginning of each semester. 

In the first lab session, the TAs delivered a brief presentation to students about the role of muons in experimental particle physics intended to orient students to the lab activities they would carry out and emphasize the broad relevance of their work. These presentations introduced the standard model of particle physics, the history of muon detection, and included a cloud chamber demonstration. In semester 2, to further communicate that muon detection is relevant work to the scientific community, ``partner particle physicists'' were brought in for each lab section. The partner physicists were members of the physics department (faculty, postdocs, and graduate students) whose research is in experimental particle physics. Some of them chose to deliver the introductory presentation instead of the TA and all partner particle physicists connected the students' future lab activities with their own and broader particle physics research. They also attended the final lab sessions for students' presentations and gave a brief presentation commenting on the students' projects and connecting them, again, to their own and broader particle physics research. In semester 2, students were also provided gloves while working directly with detector hardware in similar efforts to simulate authenticity and communicate relevance.

\subsection{Participants}
A total of 1,114 students enrolled in the course over both semesters. Students registered for one of 30 (semester 1) or 21 (semester 2) lab sections, with up to three lab sections taking place simultaneously across three different lab rooms. Prior to the start of each semester, which lab sections would be designated as SALT or PEPPER was determined by the scheduled room assignment. A single room was chosen to be the PEPPER room and all sections assigned by the registrar to that room were PEPPER labs. The remaining sections were SALT. Students did not have access to section designations when enrolling. Students were able to switch lab sections during the first week of classes as availability permitted. Across both semesters, only five students switched between a SALT and PEPPER lab. Four went from SALT to PEPPER and one student switched from PEPPER to SALT. Because switching between sections was minimal, we believe this is unlikely to have impacted any outcomes, supporting our quasi-experimental design. 

In semester 1, roughly 1/3 of the students were in a PEPPER condition lab and 2/3 were in a SALT condition lab. In semester 2, the proportion of students in a PEPPER condition lab increased to nearly 1/2. Demographic data (Table \ref{tab:demographics} were self-reported each semester through a presemester survey and a postsemester survey. We had a response rate of 64\% across both semesters. Table \ref{tab:record-counts} gives full response rates across both labs (SALT/PEPPER) and semesters. The course was overwhelmingly composed of first-year students. The majority of students were engineering majors; 11\% were physics, astronomy, or engineering physics majors.

\begin{table*}[]
\caption{Course-level demographics by semester and condition. These data come from any valid and completed presurvey or postsurvey. If a student's demographic response differed on the postsurvey from the presurvey, we used the postsurvey answer.}
\label{tab:demographics}
\begin{ruledtabular}
\begin{tabular}{lcccc}
 & \multicolumn{2}{c}{Semester 1} & \multicolumn{2}{c}{Semester 2} \\ \cline{2-3} \cline{4-5}
Student-level variables & SALT & PEPPER & SALT & PEPPER \\ \hline
All & 405 & 182 & 228 & 195 \\
Gender &      &        &      &        \\
\quad Woman                                       & 183   & 85    & 104 &  76    \\
\quad Man                                         & 212   & 93    & 110 & 106    \\
\quad Non-binary or other                         & 6     & 4     &   2 &   3    \\
\quad Did not disclose                            & 4     & 0     &  12 &  10    \\
Race or ethnicity                                 &      &        &      &        \\
\quad American Indian or Alaska Native            & 5    & 1      &   0  &   0    \\
\quad Asian                                       & 188  & 99     & 141  & 117    \\
\quad Black or African American                   & 33   & 7      &  12  &   1    \\
\quad Hispanic or Latino                          & 30   & 14     &   7  &  13    \\
\quad Middle Eastern or North African             & 8    & 7      &   4  &   3    \\
\quad Native Hawaiian or Other Pacific Islander   & 2    & 0      &   0  &   0    \\
\quad White                                       & 180  & 69     &  64  &  66    \\
\quad Some other race or ethnicity                & 8    & 2      &   0  &   3    \\
\quad Did not disclose                            & 4    & 0      &  12  &  10    \\
Class standing                                                   &      &        &      &        \\
\quad Freshman                                    & 395  & 177 &  195 & 156    \\
\quad Sophomore                                   & 3    & 3   &   22 &  30    \\
\quad Junior                                      & 5    & 1   &    4 &   2    \\
\quad Senior                                      & 1    & 1   &    2 &   4    \\
\quad Other                                       & 3    & 3   &    3 &   1    \\
\quad Did not disclose                            & 4    & 0   &   12 &  10    \\
Major                                                            &      &        &      &        \\
\quad Physics, astronomy, or engineering physics  & 27   & 7   &   55 &  31    \\
\quad Engineering                                 & 324  & 155 &  133 & 125    \\
\quad Life science or biology                     & 6    & 4   &    4 &   3    \\
\quad Other physical science                      & 26   & 6   &   18 &  22    \\
\quad Other                                       & 3    & 3   &    3 &   1    \\
\quad Did not disclose                            & 4    & 0   &   12 &  10    \\
Parent's highest level of education                            &      &        &      &        \\
\quad Did not complete high school                & 23   & 10 &    4 &   2    \\
\quad High school/GED                             & 100  & 30 &   45 &  36    \\
\quad Some college (but did not complete college) & 10   & 5  &    4 &   4    \\
\quad Associate's degree (2 year degree)          & 3    & 1  &   11 &   1    \\
\quad Bachelor's degree                           & 65   & 37 &   20 &  25    \\
\quad Master's degree                             & 110  & 44 &   67 &  61    \\
\quad Advanced graduate degree                    & 77   & 48 &   60 &  58    \\
\quad Not sure                                    & 1    & 1  &    2 &   0    \\
\quad Prefer not to disclose                      & 16   & 6  &   15 &   8   
\end{tabular}
\end{ruledtabular}
\end{table*}

\begin{table*}[]
\caption{Response rates by semester and condition. To be counted as ``finished'', students had to meet the following criteria: the student completed the survey; the student consented to participate in research; the student is at least 18 years of age; the student spent at least 30 seconds on any page; and the student provided some sort of identification (i.e. name, student ID). Data was imputed according to Subsection \ref{sec:MultipleImputation}. }
\label{tab:record-counts}
\begin{ruledtabular}
\begin{tabular}{lcccc}
 & \multicolumn{2}{c}{Semester 1} & \multicolumn{2}{c}{Semester 2} \\ \cline{2-3} \cline{4-5}
 & SALT & PEPPER & SALT & PEPPER \\ \hline
Total enrolled & 461 & 206 & 246 & 201 \\ \hline
Finished presurvey & 388 & 181 & 222 & 188 \\
Finished postsurvey & 275 & 110 & 192 & 168 \\
Finished both & 258 & 109 & 186 & 161 \\ \hline
Imputed presurvey & 17 & 1 & 6 & 7 \\
Imputed postsurvey & 130 & 72 & 36 & 27 \\ \hline
Total after imputation & 405 & 182 & 228 & 195 \\
\end{tabular}
\end{ruledtabular}
\end{table*}

\subsection{Outcome Constructs}
We measured student outcomes through five sets of survey questions given to students at the start and end of the semester. First, we measured students' experimentation skills through the Physics Lab Inventory of Critical thinking (PLIC)~\cite{walsh_quantifying_2019}, a closed-response assessment of students' critical thinking skills about a mass on a spring experiment. The instrument characterizes experimental physics-related critical thinking as how students decide what to trust (evaluating models and methods) and what to do (deciding next steps in an investigation)~\cite{walsh_assessing_2020}. Second, we measured students' attitudes about experimental physics through measures of self-efficacy, recognition, belonging, and perceived agency~\cite{feinleib_sharerotatesplit_2025}. Self-efficacy and recognition are key components of identity, a construct that, along with belonging, is hypothesized to develop systematically through engagement in research~\cite{lopatto_survey_2004, zohrabi_alaee_impact_2022, hunter_becoming_2007}. Perceived agency is a construct theoretically connected to self-efficacy~\cite{bandura_exercise_2000} and ownership~\cite{hanauer_project_2014}. We chose these constructs due to their importance in students' science identity and persistence in STEM \cite{carlone_understanding_2007, hazari_connecting_2010, godwin_identity_2016}. They are indicators of how students see themselves within their ``figured worlds'' of physics, and are therefore of particular interest for understanding the impact of these labs with regard to promoting broader participation in physics.

%See that paper for CFA, etc. It is the same student population (one of the courses in that paper is this paper's data). The items surveying these constructs were written specifically for the physics lab context. We define these constructs based on prior work adapting them to the PLIC . 
% All items ask students to score their agreement level for corresponding statements, such as ``I feel like an outsider in a physics lab'' for belonging or ``I am in control of doing interesting experiments in a physics lab'' for perceived agency.  

\subsection{\label{sec:MultipleImputation}Multiple Imputation}
Not all enrolled students completed both the presemester and postsemester survey, or completed all portions of the survey. To account for missingness of student survey responses, we used multiple imputation according to recommendations from \citeauthor{nissen_missing_2019} \cite{nissen_missing_2019} and using %. The missingness of data by outcome construct split by presemester and postsemester survey is given in Table \ref{tab:missingness-rates}. We imputed data following 
the directions in the supplemental materials decision tree \cite{Woods2021} associated with a manuscript by \citeauthor{Woods2024} \cite{Woods2024}.

We provide the missingness by construct and course condition in Table \ref{tab:missingness-rates}. We had low missingness across all presemester survey constructs. The missingness varied across the postsemester survey constructs. For semester 1, we have missingness up to 40\%, although given our sample size, this should not be a concern for our analysis \cite{graham1999performance}. For semester 2, our postsemester survey missingness was less than half of semester 1; missingness was around 15\%. 

\begin{table}[]
\caption{Percentage of missingness for survey items on the presurveys and postsurveys by condition and semester.}
\label{tab:missingness-rates}
\begin{ruledtabular}
\begin{tabular}{lcccc}
 & \multicolumn{2}{c}{Semester 1} & \multicolumn{2}{c}{Semester 2} \\ \cline{2-3} \cline{4-5}
Survey item & SALT & PEPPER & SALT & PEPPER \\ \hline
Presurvey & & & & \\
\quad PLIC          & 4.2\%  & 0.5\%  & 2.6\% & 3.6\% \\
\quad Self-efficacy & 4.4\%  & 2.2\%  & 3.5\% & 4.6\% \\
\quad Perceived     & 5.2\%  & 1.6\%  & 4.4\% & 4.6\% \\
\quad agency        &         &         &  &  \\
\quad Belonging     & 4.9\%  & 1.6\%  & 4.4\% & 6.2\% \\
\quad Recognition   & 4.4\%  & 1.1\%  & 3.1\% & 4.1\% \\
Postsurvey                        &         &         &  &  \\
\quad PLIC          & 32.1\% & 39.6\% & 15.8\% & 13.8\% \\
\quad Self-efficacy & 32.8\% & 40.1\% & 17.5\% & 14.9\% \\
\quad Perceived     & 32.6\% & 40.1\% & 17.1\% & 14.9\% \\
\quad agency        &         &         &  &  \\
\quad Belonging     & 32.6\% & 39.6\% & 17.1\% & 14.9\% \\
\quad Recognition   & 32.3\% & 39.6\% & 16.7\% & 14.4\% \\
\end{tabular}
\end{ruledtabular}
\end{table}

To ensure we could impute data, we tested if any data could predict presemester and postsemester survey completion. In line with prior studies \cite{nissen_missing_2019, feinleib_sharerotatesplit_2025}, we tested whether final letter grade predicted missingness. Using a t-test, we found that final letter grade was a statistically significant predictor of postsemester survey missingness for both semester 1 ($p<0.001$) and semester 2 ($p<0.001$). Similarly, we found that final letter grade was a statistically significant predictor of presemester survey missingness for both semester 1 ($p<0.001$) and semester 2 ($p<0.001$). These checks mean that grade is a statistically significant predictor for missingness and we can use grades to impute for missing values. We also ran logistic regressions accounting for random effects due to section and semester and again found grade was a statistically significant predictor of presemester and postsemester survey missingness (see the Supplemental Material \cite{supplementalMaterial}). 

\begin{figure}
    \centering
    \includegraphics[width=1\linewidth]{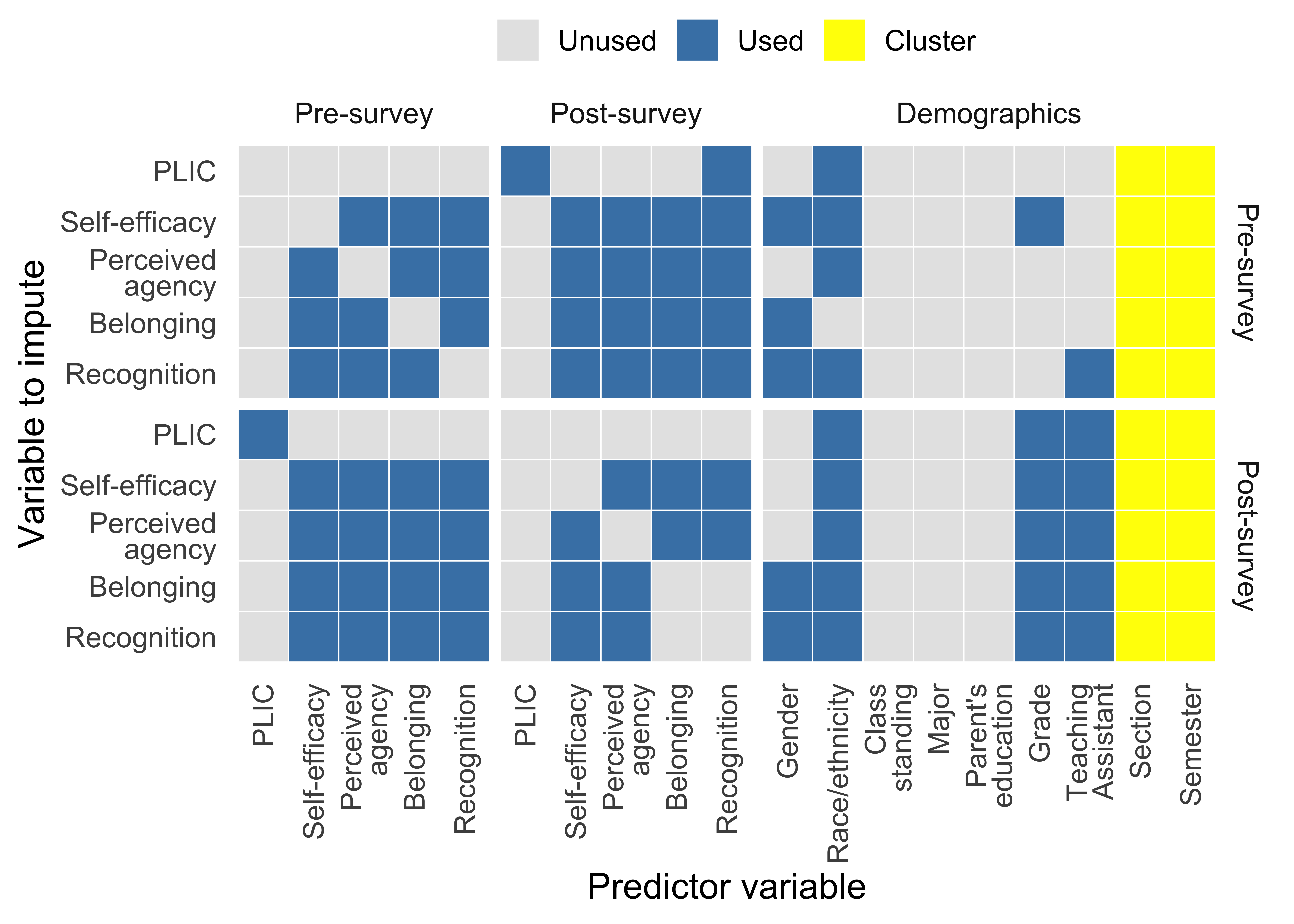}
    \caption{Predictor matrix for multiple imputation. ``Section'' and ''Semester'' were used as cluster variables for all variables. In the multiple imputation, race/ethnicity was several separate binary variables, but we only indicate in the figure whether any race/ethnicity variable was used. All options are listed in Table \ref{tab:demographics}.}
    \label{fig:predmatrix}
\end{figure}

We imputed PLIC scores and the four attitude constructs for the presemester and postsemester survey using \verb|ml.lmer| from \textit{mice} R package \cite{mice_package} for multi-level predictive mean matching. Predictive mean matching predicts values based on similar available data, which means it can handle both nonlinear and linear data well \cite{vanginkel_2020}. To improve imputation, we normalized all our data to a 0 to 1 scale. We used the \verb|quickpred| function to determine which variables to use when running our imputation for each imputed variable, as seen in Fig. \ref{fig:predmatrix}.\footnote{We also imputed grade for six students who received a ``W'' for withdrawing from the course. We imputed for grade because the \textit{MICE} package requires all variables used in multiple imputation to either be imputed or not missing. We treated ``W'' as missing because it does not map directly onto the 0--4.3 scale.}

We multiply imputed $m=50$ times, as done in a similar study \cite{feinleib_sharerotatesplit_2025}. We ensured that each imputed variable met the recommended threshold \cite{madleydowd_2019, white_2011} such that
$$ \text{FMI}/m \leq 0.01 $$
where FMI is the Fraction of Missing Information \cite{madleydowd_2019, white_2011}. We provide descriptive statistics for each variable using listwise deletion and multiple imputation in Table \ref{tab:descriptive-statistics} of Appendix \ref{sec:desc_stat_Appendix}.

\subsection{\label{sec:HLMAnalysis}Hierarchical Linear Modeling Analysis}

We used hierarchical linear modeling \cite{van_dusen_modernizing_2019, theobald_students_2018} to analyze changes in student outcomes as a result of lab condition. Hierarchical linear modeling is a type of regression that takes the nested structure of data into account. In this case, we have students nested within sections, which are nested within semesters. For this reason, we used a three-level model. To make our results easier to interpret, we normalized our attitude constructs from a 1 to 5 scale to a 0 to 1 scale. The PLIC is already scored on a 0 to 1 scale. 

For level-1 data (student level), the normalized score for a construct at the end of the semester for student $i$ in section $j$ for semester $k$, $Post_{ijk}$, are modeled in Eq. (\ref{eq:HLMLevelOne}):
\begin{align} \label{eq:HLMLevelOne}
    Post_{ijk} &= \beta_{0jk} + \beta_{1jk} Pre + r_{ijk}
\end{align}
where $\beta_{0jk}$ is the intercept term; $Pre$ is the student's normalized score on the construct at the start of the semester; and $r_{ijk}$ is the residual term.

Our level-2 data (section level) are modeled by Eq. (\ref{eq:HLMLevelTwo}),
\begin{align}  \label{eq:HLMLevelTwo}
    \beta_{0jk} &= \gamma_{00k} + \gamma_{01k} PEPPER  + u_{0jk}
\end{align}
where $\gamma_{00k}$ is the intercept term; $PEPPER$ indicates if the section was assigned the PEPPER condition; and $u_{0jk}$ is the residual term.

Finally, our level-3 data (semester level) are modeled by Eq. (\ref{eq:HLMLevelThree}),
\begin{align}  \label{eq:HLMLevelThree}
    \gamma_{00k} &= \theta_{000} + v_{00k}
\end{align}
where $\theta_{000}$ is the intercept term and $v_{00k}$ is the residual term. We check the assumptions required to use hierarchical linear modeling in Appendix \ref{sec:HLMAppendix}. 

We also considered alternate models that included additional variables (\textit{i.e.}, gender, major, first-generation status, race/ethnicity, teaching assistant) that were theorized to affect student outcomes. We compared the AIC and BIC values resulting from the analysis with data using both listwise deletion and multiple imputation. We found that the model without any of these terms consistently had the lowest or second lowest AIC or BIC. These results are included in the Supplemental Material \cite{supplementalMaterial}. Because the base model outperformed these others, we do not consider these additional terms in the model used to answer research question 1 regarding how varying broad relevance of lab content relates to student outcomes. We consider these additional variables to answer research question 2, as described in the next section.

We also evaluated whether the same outcomes would result from analysis that only evaluated physics majors. We expected the PEPPER condition was plausibly most beneficial to physics majors because the content of PEPPER labs is more relevant to professional physicists. In Appendix~\ref{sec:PhysicsMajorsAppendix}, we share the results of an analysis including only physics majors to evaluate the effect of lab condition for physics majors only.  

\subsection{\label{sec:EmmeansMethods}Estimated Marginal Means}

To answer research question 2, we used estimated marginal means to compare model-predicted outcomes across demographic groups. This analysis builds from an equity of individuality lens~\cite{rodriguez_impact_2012, van_dusen_equity_2020, burkholder_examination_2020}, which asks whether an intervention improves outcomes for demographic groups of students but without comparing outcomes between the groups. Here, we adapt this lens to evaluate whether the two lab conditions provide differential improvements for subgroups of students. Thus, we evaluated the effect of lab condition on measured demographic variables of gender, first-generation status, race/ethnicity, and student major. Estimated marginal means were calculated separately for each demographic group using the hierarchical linear model described in Sec. \ref{sec:HLMAnalysis} with added terms for the demographic group and its interaction with lab condition. We used the \verb|emmeans| function from the \textit{emmeans} package \cite{emmeans_package} to compute these values. 

%The EMM itself is the model predicted mean value for a specific demographic group and lab condition, like women in PEPPER sections, averaging over all other variables in the model.

%EMMs do not directly calculate the effects of any demographic variables, but instead provide comparisons between groups that can serve as preliminary tests for smaller datasets.

% ran it seperately and cite the emmmeans package

\section{\label{sec:Results}Results}

\begin{figure*}
    \centering
    \includegraphics[width=1\linewidth]{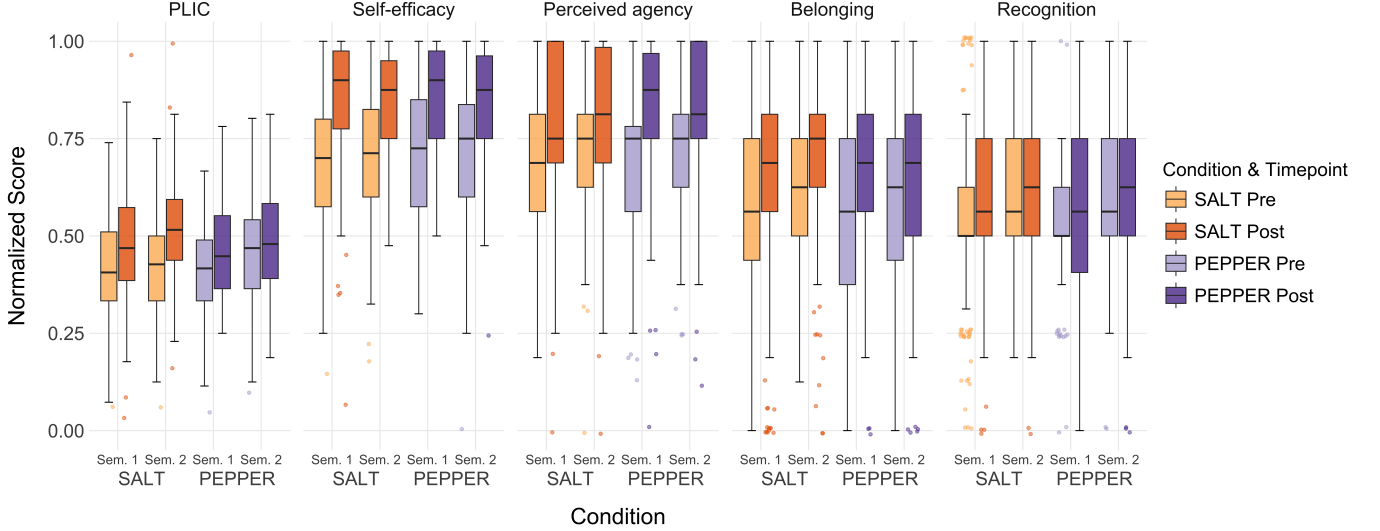}
    \caption{Box and whisker plots showing the distribution of presurvey and postsurvey scores for each construct across each semester and lab condition. The bold horizontal line is the median of the distribution, and the colored box is the inter-quartile range (IQR). Whiskers extend to 1.5 times the IQR above and below the median. Dots represent outliers outside the range of the whiskers. Data are presented here using listwise deletion.}
    \label{fig:DescStats}
\end{figure*}

\subsection{Descriptive Statistics}
Mean construct scores and standard deviations on the presurvey and postsurvey for both SALT and PEPPER conditions are shown in Fig. \ref{fig:DescStats} and reported in Table \ref{tab:descriptive-statistics} of Appendix \ref{sec:desc_stat_Appendix}. Table \ref{tab:descriptive-statistics} reports data using listwise deletion and multiple imputation, while Fig. \ref{fig:DescStats} plots data using listwise deletion. 

We observed increases in students' mean postsurvey scores for every construct (PLIC, self-efficacy, perceived agency, belonging, and recognition), across all lab conditions and semesters. We calculated effect sizes using Cohen's $d$. Using both listwise deletion and multiple imputation, the largest effect sizes consistently occur for self-efficacy ($d>0.8$) and the smallest for recognition ($0.1<d<0.3$) and belonging ($0.2<d<0.5$). Figure \ref{fig:DescStats} also indicates ceiling effects for the self-efficacy and perceived agency constructs, as seen previously with these courses in Ref. \cite{feinleib_sharerotatesplit_2025}.

% PEPPER labs go from \textit{d} = 0.97 $\pm$ 0.24 in Semester 1 to \textit{d} = 0.71 $\pm$ 0.17 in Semester 2, while SALT goes from \textit{d} = 0.93 $\pm$ 0.15 to \textit{d} = 0.78 $\pm$ 0.16. Effect sizes for belonging in PEPPER also decrease between semesters, going from \textit{d} = 0.45 $\pm$ 0.21 to \textit{d} = 0.15 $\pm$ 0.15. 

% Observing only imputed data, the largest magnitude of change (0.17 raw point increase) between pre and post-survey for any mean construct score is seen in semester 1 SALT student's self efficacy scores. Students averaged 0.69 $\pm$ 0.18 on the pre-test and increased to 0.86 $\pm$ 0.14, representing a Cohen's \textit{d} of 0.9303 $\pm$ 0.1542 (95\% CI). We saw the smallest magnitude increase(what?) in semester 1 PEPPER student's recognition scores, which shifted from 0.53 $\pm$ 0.17 on the pre-test to 0.56 $\pm$ 0.25 on the post-test for a Cohen's \textit{d} of 0.1735 $\pm$ 0.2008 (95\% CI). However, the smallest actual effect size (Cohen'd \textit{d}) was seen in semester 2 PEPPER student's belonging scores (\textit{d} = 0.1495 $\pm$ 0.1527, 95\% CI). 

%In semester one, SALT PLIC mean 0.42/0.13 -> 0.47/0.14 and PEPPER 0.4/0.13 -> 0.45/0.12. Similar magnitude  ncreases in semester two. 
%Pick out most and least improved construct scores and state increases in those here?
%\^ maybe, could just briefly talk about general trends and highlight anything notable. A benefit of giving a couple of extreme examples though would be it gives the reader a quick sense of the range of data without going through the table  

\subsection{Impact of lab conditions on student outcomes}

\begin{figure}
    \centering
    \includegraphics[width=1\linewidth]{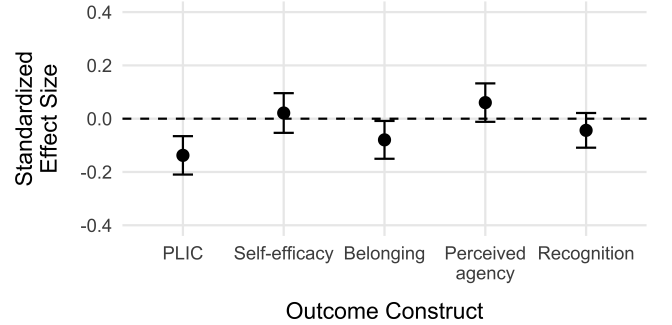}
    \caption{Results of hierarchical linear modeling of change in PLIC and attitude scores controlling for section and semester due to the PEPPER intervention. A value greater than zero indicates a positive effect on score, whereas a value less than zero indicates a negative effect on score. The horizontal dashed line indicates no observable effect on score. The error bars represent the standard error of the effect size. There were no statistically significant effects. Plot was made using multiple imputation for missing data.}
    \label{fig:hlm_results}
\end{figure}

We answer our first research question---about the overall effects of lab condition on student outcomes---using a hierarchical linear model. Table \ref{tab:HLMresult} and Fig. \ref{fig:hlm_results} report the effect sizes, which correspond to the coefficient $\gamma_{01k}$ in Eq. (\ref{eq:HLMLevelTwo}), standardized to represent the number of standard deviations by which scores shifted. A positive effect indicates that the expected score on the postsurvey for a student in the PEPPER lab condition is higher than that for a student with the same presurvey score in the SALT lab condition. A negative effect indicates the opposite.

% Before imputing missing data, 

Using listwise deletion, we observed no statistically significant effects ($p > 0.05$) for any construct except the PLIC ($\gamma_{01k}=-0.1820 \pm 0.0755$, $p = 0.016$). Using multiple imputation, this effect was not statistically significant ($-0.1376 \pm 0.0721$, $p= 0.057$); in fact, no effect sizes were statistically significant using multiple imputation. 

% Before imputing, the PEPPER condition had a negative effect on all constructs with the exception of perceived agency. After imputation, effects for 3 out of 5 constructs were negative while self-efficacy and perceived agency were positive. Effect sizes ranged from -0.1820 $\pm$ 0.0755 (\textit{p} = 0.016) for PLIC score without imputation to 0.0731 $\pm$ 0.0733 (\textit{p} = 0.319) for perceived agency without imputation. 

\begin{table}[]
\begin{ruledtabular}
\caption{Results from hierarchical linear modeling for change in outcome construct scores. The table includes the standardized effect sizes, standard errors, and $p$ values in parentheses.}
\label{tab:HLMresult}
\begin{tabular}{lc}
Construct & Effect \\ \hline
Using listwise deletion & \\
\quad PLIC & -0.1820 $\pm$ 0.0755 (0.016) \\
\quad Self-efficacy & -0.0104 $\pm$ 0.0744 (0.889) \\
\quad Belonging & -0.1157 $\pm$ 0.0698 (0.098) \\
\quad Perceived agency & 0.0731 $\pm$ 0.0733 (0.319) \\
\quad Recognition & -0.0612 $\pm$ 0.0665 (0.358) \\
Using multiple imputation & \\
\quad PLIC & -0.1376 $\pm$ 0.0721 (0.057) \\
\quad Self-efficacy & 0.0213 $\pm$ 0.0745 (0.775) \\
\quad Belonging & -0.0795 $\pm$ 0.0710 (0.263) \\
\quad Perceived agency & 0.0603 $\pm$ 0.0721 (0.403) \\
\quad Recognition & -0.0438 $\pm$ 0.0651 (0.502)
\end{tabular} 
\end{ruledtabular}
\end{table}

In Appendix \ref{sec:PhysicsMajorsAppendix}, we also evaluate the impact of lab condition on physics majors only. we observed no statistically significant effects ($p > 0.05$) for any construct except the PLIC; physics majors in PEPPER scored lower than physics majors in SALT on the PLIC ($\gamma_{01k} = -0.4565 \pm 0.2032$, $p=0.027$ using multiple imputation).

\subsection{Differential impacts across demographic groups}

\begin{figure*}
    \centering
    \includegraphics[width=1\linewidth]{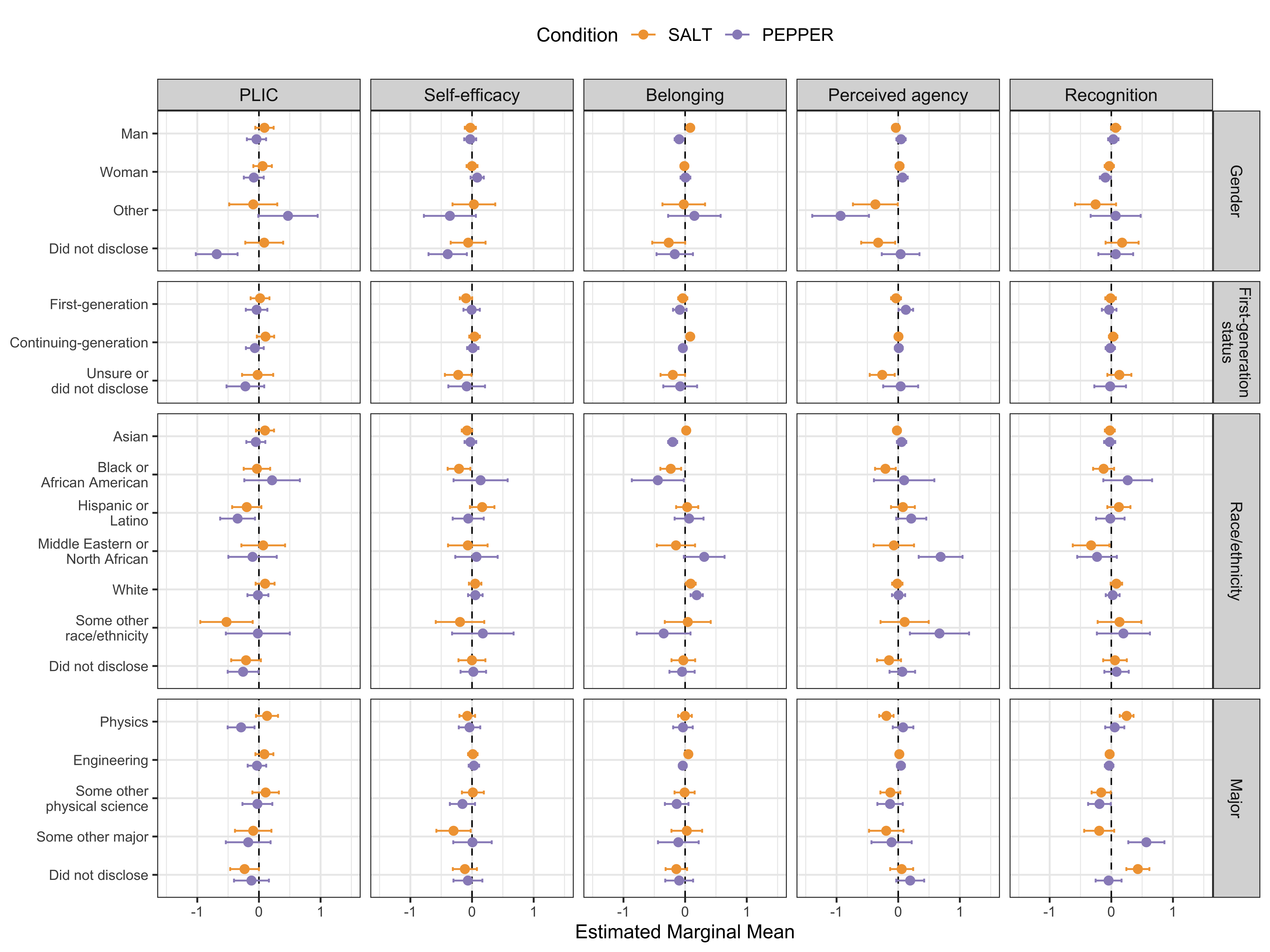}
    \caption{Estimated marginal means (EMMs) for construct scores of different demographics of students. EMMs are standardized as standard deviations relative to the overall postsurvey mean distribution. The center dashed line represents the overall postsurvey mean for that construct. Error bars represent the standard error of the EMM. Plot was made using multiple imputation for missing data.}
    \label{fig:emmeans_figure}
\end{figure*}

We answer our second research question---about potential differential effects of lab condition on different student groups---using estimated marginal means (EMMs) for each outcome construct and demographic variable (Fig. \ref{fig:emmeans_figure}). EMMs are derived from the same hierarchical linear model presented in the previous section, but with interaction terms between demographic variables and the lab condition added. Although the interaction terms directly capture differential effects by demographic group, we present EMMs to facilitate interpretability and to emphasize the exploratory nature of the analysis, as we do not have sufficient statistical power to meaningfully interpret the effect sizes on the interaction terms. Further, whereas interaction coefficients are always in reference to some baseline group, making cross-group comparison inconvenient, EMMs are reported on a common scale relative to the overall postsurvey mean for a given construct. This presentation is more appropriate for exploratory analysis where the goal is visual identification of patterns rather than formal hypothesis testing. A larger study designed to investigate demographic differences would need to collect far more data. Still, this exploratory analysis provides an important check on whether there are potential differential responses by demographic group and serves to inform future study designs.

We report EMMs in standard deviations relative to the overall postsurvey mean for that construct. Error bars represent the standard error of the EMM. An EMM greater than zero indicates that a demographic group is predicted to score above the overall mean, and an EMM less than zero indicates that a demographic group is predicted to score below. %Critically, these comparisons are post-hoc and descriptive in nature. 

%incorporate these variables directly into the hierarchical linear model, enabling formal tests of differential effects across groups. 

% other reasons we're doing emmeans - focus on visual patterns rather than hypothesis testing? interpretability (no reference group) so patterns would be more readily apparent? EMMs derived from HLM model with interaction terms

We see no meaningful trends in differences in EMMs between demographic groups for any construct in Fig. \ref{fig:emmeans_figure}. For a majority of demographic groups, EMMs are clustered near zero across both SALT and PEPPER conditions, suggesting no strong pattern of differential effect by gender, first generation status, race/ethnicity, or major. Demographic groups that show greater variability in EMMs also have larger standard errors, reflecting increased uncertainty in the estimates due to small sample sizes rather than statistically meaningful differences.  

As a concrete example, consider the top-left box showing PLIC EMMs of different gender demographic groups. Throughout the plot, EMMs for the PEPPER condition are shown in purple, and EMMs for the SALT condition are shown in orange. The postsurvey PLIC EMM for men in SALT sections is slightly above zero, indicating predicted scores slightly above the overall mean. The EMM for PEPPER sections is slightly below zero, indicating predicted scores slightly below the overall mean. However, we see that these two EMMs are not meaningfully different from each other. The same pattern exists for the postsurvey PLIC EMMs of women in SALT and PEPPER. We do see a visual difference between SALT and PEPPER EMMs for the ``Did not disclose'' gender group, but the standard error for this group is large due to a much smaller sample size.

%All following results remain unchanged regardless of method used to handle missing data - MI or listwise deletion. 
%Based on HLM analysis, we did not observe a significant effect of the PEPPER lab section on PLIC and outcome constructs (list...) compared to SALT labs. 
%Effect sizes reported in table VI
%P values not significant for anything at 0.05 level
%Overall PLIC effect -0.1820/0.0755
%Again, quote highest and lowest effect sizes in paragraph? 

%new paragraph for subsets/AIC/BIC
%Repeated HLM analysis for subset of physics, engineering physics, and astronomy majors. 
%Quote highest and lowest effect sizes, compare to the non subsetted results above?
%appendix B
%table VII, fig. 
%We also tested models that accounted for gender, major, first-generation status, race/ethnicity, and lab section teaching assistant. 
%These models did not improve over the base model. We determined this by comparing the AIC/BIC values for the base model and model with additional factors. 

\section{\label{sec:Discussion}Discussion and Conclusions}

In this study, we sought to disentangle the effect of broad relevance (a core component of CUREs) on student outcomes. We found no statistically significant difference in posttest scores between students in the CURE-like PEPPER lab compared to the experimentation-based SALT lab on five measured constructs: experimental critical thinking skills, self-efficacy, perceived agency, belonging, and recognition. We infer that systematically varying lab activities to increase their relevance to contemporary physics research did not improve student gains in these critical domains, which are strongly linked to science identity and persistence \cite{carlone_understanding_2007, hazari_connecting_2010, godwin_identity_2016}. Students similarly improved their critical thinking and attitudes regardless of whether the physics was ``cutting edge'' or not. Moreover, we observe no evidence of either lab condition disproportionately benefiting certain demographic groups over others, although more data is needed to more definitively claim that the benefits from each lab condition impact students equitably. 

The results of this study have several important implications for understanding CURE and CURE-like models in physics education, particularly for institutional contexts where the incorporation of novel research into the lab curriculum may be untenable. For one, the present study shows that altering the context of introductory labs to make them more broadly relevant does not automatically boost gains in experimental critical thinking skills, self-efficacy, belonging, agency, and recognition beyond what is achievable with other experimentation-based labs. Furthermore, although this study did not implement a full CURE, our findings align with a growing body of evidence that has begun to challenge the idea that CUREs are uniquely positioned to achieve superior student outcomes \cite{rowland_we_2016, ballen_discovery_2018, hebert2021open, lansverk2020comparing}. It remains an open question as to whether implementing a well-designed experimentation-based lab in physics may be able to produce comparable student gains to a CURE. Together with prior work, however, the current study supports the idea that broad relevance may not be essential to achieving a number of valued student outcomes. It also points to the need to understand how other forms of relevance (\textit{e.g.}, personal, societal \cite{stuckey2013meaning}) shape students' lab experiences. More precisely explicating how these constructs are related to engaging students in broadly relevant work is particularly important, as the role of broad relevance may be confounded when different lab formats also tap into these other dimensions of relevance.

% Broad relevance is the element that most strongly distinguishes CUREs from other non-traditional, skills-based labs. It is also the component that makes CUREs most difficult to implement because it requires departments to continually develop research projects suitable for large cohorts of undergraduates. Because CUREs engage students in novel research, they are often regarded as the gold-standard lab experience for students.   

Although our sample size was not large enough to make definitive claims about differential effects across demographic groups, we observed no evidence that the benefits of either lab condition were concentrated among particular demographic groups. This tentatively suggests that well-designed experimentation-based labs may be capable of supporting persistence in a broad range of students, which is a critical consideration for departments seeking equitable and sustainable approaches to lab reform. Unlike the discussion above, however, we cannot draw on the larger body of literature to strengthen this claim, as the equity implications of experimentation-based labs relative to CUREs were not explored in those studies. A more definitive result will require larger samples across demographic groups. CUREs have been highlighted as an effective means for supporting students from underrepresented groups in science \cite{bangera_course-based_2014, werth_impacts_2022, estrada2016improving}, so better understanding whether other lab formats equitably support students will be an essential line of inquiry moving forward.

While the results presented in this paper may appear to contradict the idea that more authentic lab experiences boost psychosocial outcomes for students, a closer examination of the literature on authenticity offers one explanation. Broad relevance is emphasized in CUREs as providing students with a more ``real'' or ``authentic'' experience \cite{auchincloss_assessment_2014, rowland_we_2016}; but authenticity, like relevance, is notoriously ill-defined in the literature \cite{rowland_we_2016, hutchison2008epistemological, schriebl2023modelling}. Some authors emphasize that authentic research is characterized by both ``process'' and ``product,'' meaning that students must contribute to publishable research products for it to be regarded as truly authentic \cite{weaver_inquiry-based_2008, auchincloss_assessment_2014, spell_redefining_2014}. In this view, which aligns more with the defining features of a CURE, broad relevance is critical. However, other authors argue that authenticity is more strongly rooted in the \textit{process} part of science, including students' ability to take ownership of a project and engage in scientific reasoning, and that this process is more important to supporting student perceptions of authenticity \cite{barab_doing_2001, peffer_science_2015}. This perspective aligns with constructivist views of authenticity as an emergent, student-centered quality \cite{rahm_value_2003, rivera_maulucci_urban_2014}. Rather than being a fixed property, authenticity in this view depends on context and on students' own perceptions of their experience. If authenticity is an emergent property of student engagement rather than a feature of the task itself, boosting the disciplinary relevance of the lab activities may not reliably produce a more authentic experience in the eyes of students. Instead, students in both SALT and PEPPER may have perceived their work as similarly authentic due to the analogous structural and process aspects of the course, such as being able to choose and pursue their own research question. Whether, how, and why the students in the two lab sections perceive their experiences as real, relevant, and authentic (or not) are open questions that we plan to investigate in a future study.

%For example, \citeauthor{martin_authentic_1990} advocates for approaching authenticity as it relates to the learner, determining what would support students best in their particular context, ways of thinking, and purpose \cite{martin_authentic_1990}. In this view it is not viable or useful for authenticity to be determined a priori by disciplinary experts. Rather, this view emphasizes the design of experiences that engage students in ways that best support their development as physicists and retention in the discipline. 

One interpretation for the results presented here is that broad relevance may simply not affect the outcomes measured by our five assessment constructs, but may impact other student outcomes not measured here. It is also possible that students in the PEPPER lab \textit{did} perceive their experience as more authentic than students in the SALT lab, but that it was not strong enough to see an effect. This study and others \cite{corwin2018need, ballen_discovery_2018} may not have provided sufficient contrast in broad relevance between lab conditions to produce a measurable effect. However, \citeauthor{beck2023can} raise doubt about whether such a threshold of relevance meaningfully exists or whether it could even be detected by measurement instruments currently favored in the CURE literature \cite{beck2023can}. 

Another alternative explanation for the data is that broad relevance \textit{does} reliably promote improved outcomes, but that those benefits depend on other social and contextual factors that are difficult to replicate in a large introductory course. This would align with \citeauthor{zohrabi_alaee_impact_2022}'s previous findings in the context of undergraduate research, which indicated that students’ engaging in novel knowledge production primarily served to foster a sense of meaningful contribution to their research groups. This feeling that their contributions were valued by their group, in turn, strengthened their physics identity \cite{zohrabi_alaee_impact_2022}. In a large lab class like PEPPER, that mechanism is largely absent. Students do not have a tight-knit research group whose members can recognize and reinforce the value of their contributions. Doing broadly relevant work in the context of a large lab class may therefore not carry the same impact as a small group.

It is worth reiterating that our study did not implement a full CURE, and the conclusions above reflect a synthesis of evidence across studies comparing CUREs and experimentation-based labs, as well as our own study comparing a CURE-like lab to an experimentation-based lab. Future work is still needed to more fully characterize the impact of broad relevance in labs and how CURE outcomes compare to those in other experimentation-based labs. Along these lines, CUREs notably offer several benefits that experimentation-based labs are unable to provide: Students in a CURE may be listed as authors on a publication related to their work in the course, which can support future employment or applications to graduate school. In this sense, CUREs promote the democratization of scientific research in a way that other types of labs cannot. However, if the primary goal of that democratization is to allow more students to experience authentic science in ways that support persistence and growth as scientists, then present evidence does not preclude the possibility that experimentation-based labs are sufficient to achieve this goal.

Looking ahead, this work underscores the need to go beyond testing benefits of different lab designs against traditional verification labs. Most work on introductory labs to date has focused on demonstrating the benefits of shifting away from traditional verification lab formats, rather than the relative strengths and weaknesses of different non-traditional lab designs. As a result, there remains limited evidence distinguishing which specific elements of non-traditional labs are most strongly associated with particular student outcomes. Yet these distinctions become critical when departments must weigh the benefits of different approaches to lab reform against the substantial variation in resources and institutional support required to implement them. Understanding how and for whom different curricula are expected to improve learning outcomes is essential for making sustainable and equitable decisions.

\begin{acknowledgments}
We are incredibly grateful to Spencer Axani for advice about the detectors. We thank Matt Thomas, Kenneth Tyler Wilcox, and the Cornell Statistical Consulting Unit for their help with data analysis, as well as Alexandra Werth and the Cornell Discipline-based Education Research community for feedback on this manuscript. We appreciate the support of Cristina Schlesier and the TAs and students who participated in the courses. This material is based upon work supported by the National Science Foundation Grant No. DGE-2139899 and PHY-2310035. We also acknowledge support from the Cornell Nexus Scholars program.

E.M. and M.V. led the writing of the manuscript. E.M. and M.D. led the data analysis. E.M., X.C., M.L-M., S.P., H.S., P.W., and N.G.H. led the development of the PEPPER lab. N.G.H. led the data collection and implementation of the course materials. All authors reviewed and edited the manuscript.
\end{acknowledgments}

\section*{Data Availability}
The data that support the findings of this article are openly available \cite{supplementalMaterial}.

%add like a sentence here abt that they exist and are in table whatever
\appendix
\section{Descriptive Statistics}\label{sec:desc_stat_Appendix}
See Table \ref{tab:descriptive-statistics} for the descriptive statistics of measured constructs using listwise deletion and using multiple imputation.   

\begin{table*}[]
\caption{Descriptive statistics (mean and standard deviation) using listwise deletion and multiple imputation for each condition and semester.}
\label{tab:descriptive-statistics}
\begin{ruledtabular}
\begin{tabular}{lcccc}
 & \multicolumn{2}{c}{Semester 1} & \multicolumn{2}{c}{Semester 2} \\ \cline{2-3} \cline{4-5}
Survey item & SALT & PEPPER & SALT & PEPPER \\
\hline
Using listwise deletion               &             &             &  &  \\
Pre-survey                            &             &             &  &  \\
\quad PLIC             & $0.42 \pm 0.13$ & $0.40 \pm 0.13$ & $0.43 \pm 0.13$ & $0.46 \pm 0.13$ \\
\quad Self-efficacy    & $0.69 \pm 0.17$ & $0.71 \pm 0.17$ & $0.71 \pm 0.16$ & $0.72 \pm 0.17$ \\
\quad Perceived agency & $0.70 \pm 0.19$ & $0.68 \pm 0.21$ & $0.73 \pm 0.17$ & $0.72 \pm 0.17$ \\
\quad Belonging        & $0.57 \pm 0.21$ & $0.55 \pm 0.22$ & $0.61 \pm 0.20$ & $0.60 \pm 0.21$ \\
\quad Recognition      & $0.53 \pm 0.19$ & $0.52 \pm 0.16$ & $0.60 \pm 0.17$ & $0.58 \pm 0.18$ \\
Post-survey                           &             &             &  &  \\
\quad PLIC             & $0.47 \pm 0.14$ & $0.45 \pm 0.12$ & $0.51 \pm 0.14$ & $0.49 \pm 0.12$ \\
\quad Self-efficacy    & $0.86 \pm 0.14$ & $0.87 \pm 0.12$ & $0.85 \pm 0.13$ & $0.85 \pm 0.13$ \\
\quad Perceived agency & $0.80 \pm 0.17$ & $0.81 \pm 0.20$ & $0.80 \pm 0.18$ & $0.82 \pm 0.17$ \\
\quad Belonging        & $0.67 \pm 0.24$ & $0.66 \pm 0.23$ & $0.70 \pm 0.21$ & $0.65 \pm 0.24$ \\
\quad Recognition      & $0.59 \pm 0.22$ & $0.56 \pm 0.26$ & $0.65 \pm 0.22$ & $0.64 \pm 0.21$ \\
Using multiple imputation             &             &             &  &  \\
Pre-survey                            &             &             &  &  \\
\quad PLIC             & $0.41 \pm 0.13$ & $0.41 \pm 0.13$ & $0.42 \pm 0.13$ & $0.45 \pm 0.13$ \\
\quad Self-efficacy    & $0.69 \pm 0.18$ & $0.71 \pm 0.17$ & $0.71 \pm 0.17$ & $0.73 \pm 0.16$ \\
\quad Perceived agency & $0.71 \pm 0.19$ & $0.71 \pm 0.19$ & $0.72 \pm 0.17$ & $0.73 \pm 0.16$ \\
\quad Belonging        & $0.57 \pm 0.22$ & $0.55 \pm 0.22$ & $0.60 \pm 0.22$ & $0.61 \pm 0.21$ \\
\quad Recognition      & $0.53 \pm 0.18$ & $0.53 \pm 0.17$ & $0.59 \pm 0.18$ & $0.58 \pm 0.18$ \\
Post-survey                           &             &             &  &  \\
\quad PLIC             & $0.47 \pm 0.14$ & $0.46 \pm 0.13$ & $0.51 \pm 0.14$ & $0.49 \pm 0.12$ \\
\quad Self-efficacy    & $0.85 \pm 0.14$ & $0.87 \pm 0.12$ & $0.84 \pm 0.13$ & $0.85 \pm 0.13$ \\
\quad Perceived agency & $0.80 \pm 0.18$ & $0.80 \pm 0.19$ & $0.80 \pm 0.18$ & $0.82 \pm 0.17$ \\
\quad Belonging        & $0.66 \pm 0.24$ & $0.66 \pm 0.22$ & $0.69 \pm 0.21$ & $0.65 \pm 0.24$ \\
\quad Recognition      & $0.59 \pm 0.22$ & $0.57 \pm 0.25$ & $0.64 \pm 0.23$ & $0.64 \pm 0.21$
\end{tabular} 
\end{ruledtabular}
\end{table*}

\section{Assumptions for Hierarchical Linear Modeling}\label{sec:HLMAppendix}
In this appendix we check if our data meets assumptions required for hierarchical linear modeling according to prior recommendations \cite{van_dusen_modernizing_2019}: the assumptions of linearity, homogeneity of variance, and normality. We checked these assumptions for all five hierarchical linear models that we ran, \textit{i.e.}, one model for each of the five constructs according to the model in Sec. \ref{sec:HLMAnalysis}. 

To check these assumptions, we used data pooled from multiple imputation rather than data using listwise deletion. We do not know of any recommendations for or against using pooled data rather than data using listwise deletion, but we chose to use pooled data in line with other studies \cite{nissen_investigating_2021, feinleib_sharerotatesplit_2025}. We used \verb|lmer| from \textsc{lme4} in R for our hierarchical linear modeling \cite{lme4Package}. 

\begin{figure*}
    \centering
    \includegraphics[width=1\linewidth]{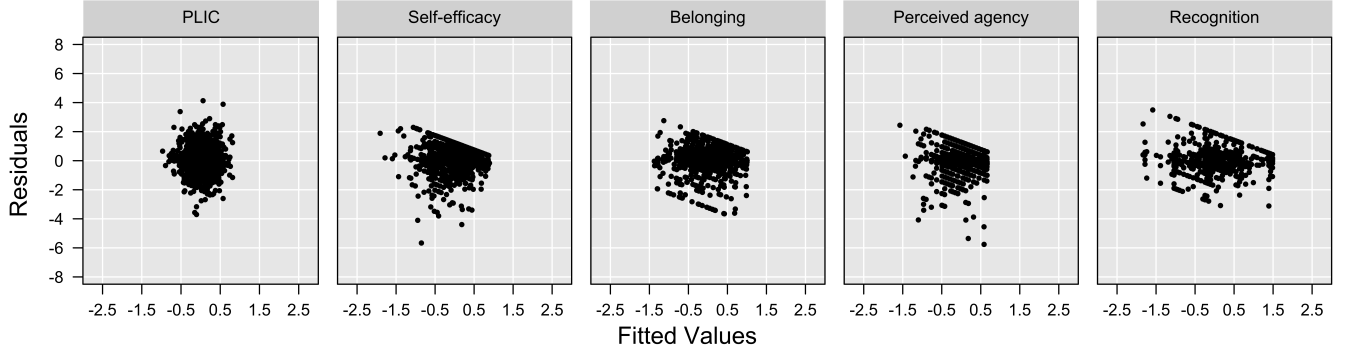}
    \caption{A visual check for hierarchical linear modeling’s assumption of linearity for all five constructs. These plots display the residuals vs fitted values for each construct across both semesters of data.}
    \label{fig:hlm_linearity}
\end{figure*}

\begin{figure*}
    \centering
    \includegraphics[width=1\linewidth]{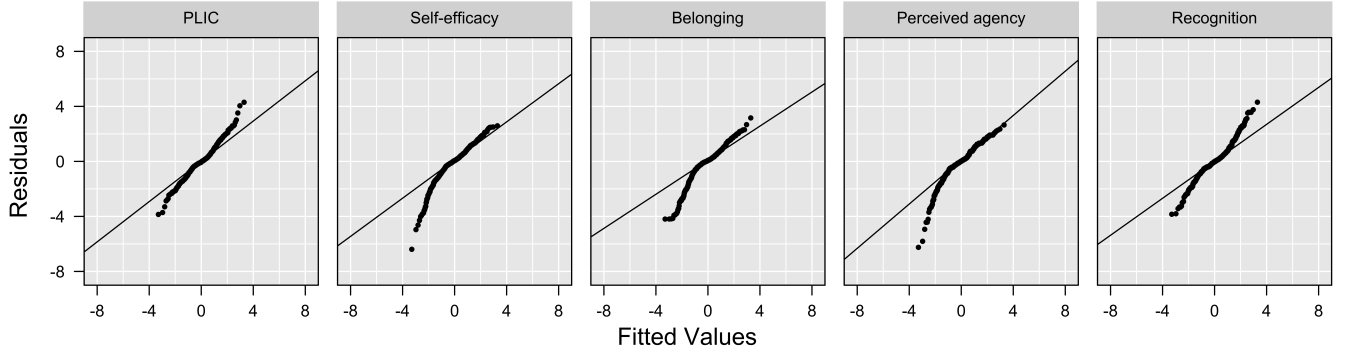}
    \caption{A visual check for hierarchical linear modeling’s assumption of normality for all five constructs. These plots display the residuals vs fitted values for each construct across both semesters of data.}
    \label{fig:hlm_normality}
\end{figure*}

A visual check for the assumption of linearity is provided in Fig. \ref{fig:hlm_linearity}. While do see ceiling effects for the student attitude constructs and floor effects for the belonging and recognition effects, we do not see large trends beyond this. We checked the assumption of homogeneity of variance using ANOVA. We found no statistically significant differences ($p>0.146$), indicating our data met this assumption. A visual check for the assumption of normality is provided in Fig. \ref{fig:hlm_normality}. These data do trail off from normality at the edges for the student attitude constructs, however, this typically does not affect statistical significance \cite{Lumley_2002, walsh_skills-focused_2022}. 

\section{CURE Effects on Physics Students}\label{sec:PhysicsMajorsAppendix}
In this section we narrow our analysis specifically to physics-related majors. Specifically, this includes students majoring in physics, astronomy, and engineering physics. The total number of physics students is low ($N_{\text{SALT}} = 82, N_{\text{PEPPER}} = 38)$ compared to our full study ($N_{\text{SALT}} = 633, N_{\text{PEPPER}} = 377)$, however, this provides additional insight for CUREs as a form of instruction specifically for physics-related majors, who have different instructional goals from other majors (\textit{e.g.}, engineering, physical sciences, etc.).

We use the same model and imputation described in Subsections \ref{sec:MultipleImputation} and \ref{sec:HLMAnalysis}, only we narrow our data set to physics and physics-related majors. We use the same imputed data set and hierarchical linear modeling equations as described in Equations \ref{eq:HLMLevelOne}, \ref{eq:HLMLevelTwo}, and \ref{eq:HLMLevelThree}. We do not recreate Tables \ref{tab:demographics}, \ref{tab:record-counts}, \ref{tab:missingness-rates}, and \ref{tab:descriptive-statistics} here. They are available in the Supplemental Material \cite{supplementalMaterial}.

The results using listwise deletion and multiple imputation are available in Table \ref{tab:HLMresultPhysicsMajors}. The results using multiple imputation are plotted in Fig. \ref{fig:hlm_results_physics}. Effect sizes remain not statistically significant ($p > 0.05$), with the exception of the PLIC score. Using both listwise deletion and multiple imputation, we see a statistically significant ($p < 0.05$) negative effect. Using multiple imputation, the effect size of the PLIC is $-0.04565 \pm 0.2032$.

% For other constructs, compared to effect sizes for all students reported in Table \ref{tab:HLMresult}, physics majors see similar effects for self-efficacy and belonging. The effect size for perceived agency is larger for physics majors, while the effect size for recognition is lower (more negative). 

\begin{table}[]
\begin{ruledtabular}
\caption{Results from hierarchical linear modeling for change in outcome construct scores for physics and physics-related majors. The table includes the standardized effect sizes, standard errors, and $p$ values in parentheses.}
\label{tab:HLMresultPhysicsMajors}
\begin{tabular}{lc}
Construct & Effect \\ \hline
Using listwise deletion & \\
\quad PLIC & -0.5240 $\pm$ 0.2066 (0.013) \\
\quad Self-efficacy & -0.0433 $\pm$ 0.2142 (0.840) \\
\quad Belonging & -0.0701 $\pm$ 0.1979 (0.724) \\
\quad Perceived agency & 0.2017 $\pm$ 0.1922 (0.297) \\
\quad Recognition & -0.3025 $\pm$ 0.1821 (0.100) \\
Using multiple imputation & \\
\quad PLIC & -0.4565 $\pm$ 0.2032 (0.027) \\
\quad Self-efficacy & 0.0113 $\pm$ 0.2169 (0.959) \\
\quad Belonging & -0.0389 $\pm$ 0.1882 (0.836) \\
\quad Perceived agency & 0.2286 $\pm$ 0.1970 (0.249) \\
\quad Recognition & -0.2167 $\pm$ 0.1838 (0.241)
\end{tabular} 
\end{ruledtabular}
\end{table}

\begin{figure}
    \centering
    \includegraphics[width=1\linewidth]{Figures/PEPPER_Effect_Physics_Impute.png}
    \caption{Results of hierarchical linear modeling of change in PLIC and attitude scores for physics majors controlling for section and semester due to the PEPPER intervention. A value greater than zero indicates a positive effect on score, whereas a value less than zero indicates a negative effect on score. The horizontal dashed line indicates no observable effect on score. The error bars represent the standard error of the effect size. The asterisk denotes statistical significance where $\ast$ indicates $p < 0.05$. Plot was made using multiple imputation for missing data.}
    \label{fig:hlm_results_physics}
\end{figure}

\bibliography{apssamp.bib, references.bib}

\end{document}